# Dynamic Critical Behavior of the Heisenberg Model with Strong Easy Plane Anisotropy

V. A. Mutailamov and A. K. Murtazaev

*Institute of Physics, Dagestan Scientific Center, Russian Academy of Sciences, Makhachkala, 367003 Russia*

*e-mail: vadim.mut@mail.ru*

**Abstract** - The dynamic critical behavior of the Heisenberg model with a strong anisotropy of the exchange constant in the *z* direction is investigated. The main features of the time evolution of this model are revealed. The static and dynamic critical behavior of planar magnetic models is shown to be described well by the Heisenberg model with strong easy plane anisotropy.

Metal magnetic superlattices are of great interest in modern condensed matter physics. The ability of external control of such properties of superlattices as magnetization, interlayer exchange interaction, magnetoresistance, and other features of these materials makes them unique objects for applications and theo retical research [1–3].

The experimental investigations of such systems meet considerable difficulties; that is why they have been recently studied by the computational physics methods. For example, the statistical critical behavior of Fe/V magnetic superlattices was investigated in [4, 5], where the statistical critical indices were determined and their dependence on the correlation between the intralayer and interlayer exchange interactions was analyzed.

The dynamic critical behavior of magnetic superlattices, which is absolutely unstudied, is no less interesting. The study of the magnetic critical dynamics is a rather complex problem. The investigation of the superlattice critical dynamics involves a number of additional difficulties. In this case, the computational physics methods also face rather serious problems. Nevertheless, the numerical experiment has become an efficient instru ment for such studies.

In this paper, we attempted to approve a new technique for studying the critical dynamics of the Fe/V magnetic superlattice. Many iron–vanadium superlattices can be described by the Hamiltonian [4, 5]

$$H = -J_{\parallel} \frac{1}{2}\sum_{i,j}\left(S_i^x S_j^x + S_i^y S_j^y\right) - J_{\perp}\frac{1}{2}\sum_{i,k}\left(S_i^x S_k^x + S_i^y S_k^y\right), \qquad (1)$$

where the first sum takes into account the exchange interaction of each magnetic atom with the nearest neighbor within the layer, and the second term considers the interaction with the atoms of the neighboring layers through the nonmagnetic interlayer; $S_i^{x,y}$ are the projections of the spin localized at site *i*. The form of the Hamiltonian indicates that the Fe/V model is in fact a kind of the *XY* model.

Generally, the spin molecular dynamics method is used as a numerical method for studying the magnetic system spin dynamics. It is based on solving the equations of spin movement in the local magnetic field [6]:

$$\frac{d\vec{S}_i}{dt} = \left[\vec{S}_i \times \vec{h}_{loc}^i\right], \qquad (2)$$

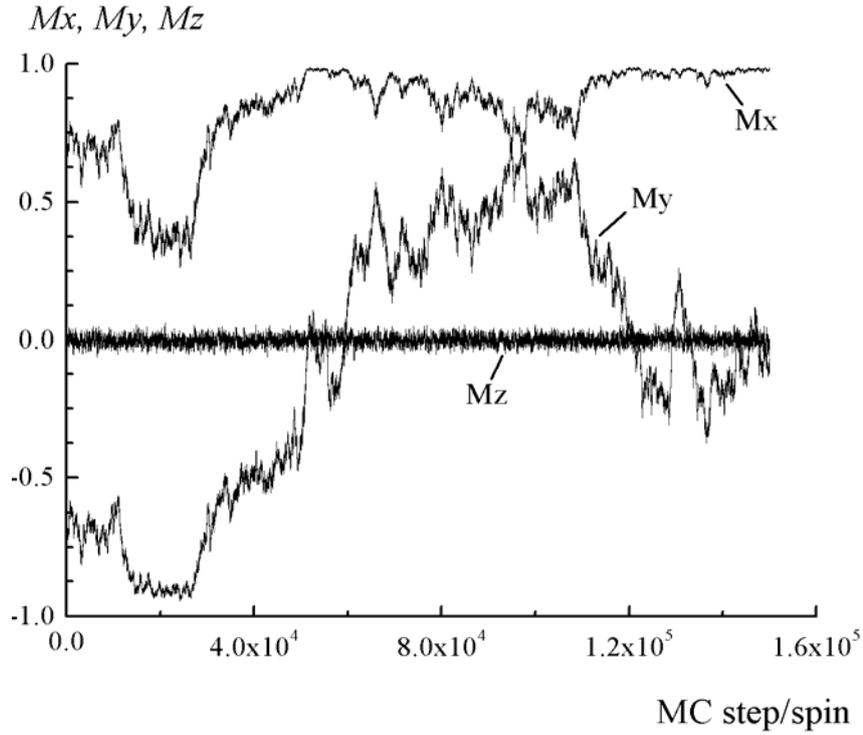

**Fig. 1.** Relaxation dependences of the magnetization vector projections $M_x$, $M_y$, and $M_z$.

where $\vec{h}_{loc}^{i}$ is the local magnetic field, which acts on the spin $\vec{S}_i$ and is determined by the Hamiltonian of the system. However, this method cannot be applied directly in the case under consideration because of the difference in the spatial spin dimensions in (1) and (2).

In this context, a special technique was proposed in [7] to investigate such systems. It is based on the use of the 3D Heisenberg model with a strong anisotropy of the exchange constant in the $z$ direction. The hamiltonian in this model can be represented as

$$H = -\frac{1}{2}\sum_{i,j}\left(J^{x}S_{i}^{x}S_{j}^{x} + J^{y}S_{i}^{y}S_{j}^{y} + J^{z}S_{i}^{z}S_{j}^{z}\right), \quad J^{x} = J^{y}, \quad J^{z} = 0. \tag{3}$$

The absence of interaction between the spin $z$ projections leads to a strong anisotropy in the $xy$ plane. Then, the critical dynamics of model (3) was investigated using the approach based on the combined use of the Monte Carlo and molecular dynamics methods [8–10].

The relaxation dependence of the magnetization vector components, obtained by the Monte Carlo method for a system including 64 spins, is shown in Fig. 1. The state of the system corresponds to the critical point $T = T_c$. As can be seen in Fig. 1, the $z$ component remains close to zero during the entire observation period. This fact confirms the suggestion that the behavior of the model studied is similar to that of the classical 3D XY model. Another confirmation of this similarity is that the statistical critical indices for the Heisenberg model with easy plane anisotropy, obtained in the numerical experiment, coincide with the similar indices of the *XY* model, although the critical temperatures in these models are different [7].

Figure 2 demonstrates the time dependences of the magnetization vector projections, obtained by the molecular dynamics method for the system including 64 spins. Apparently, the $z$ projection of the magneti_ zation vector remains a close-to-zero constant in time, and the vector itself rotates around the $z$ axis. Such behavior of magnetization vector is also typical of *XY* model. The deviation of the magnetization vector from the $xy$ plane is determined by the spin configuration (obtained within the Monte Carlo method) by the moment of applying the molecular dynamics method.

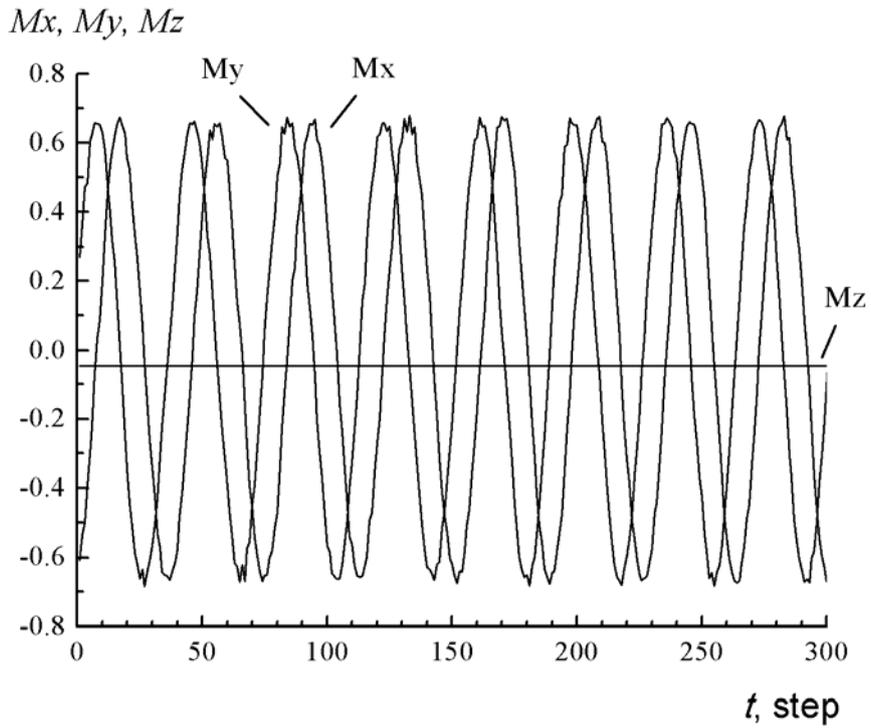

**Fig. 2.** Time dependences of the magnetization vector projections $M_x$, $M_y$, and $M_z$.

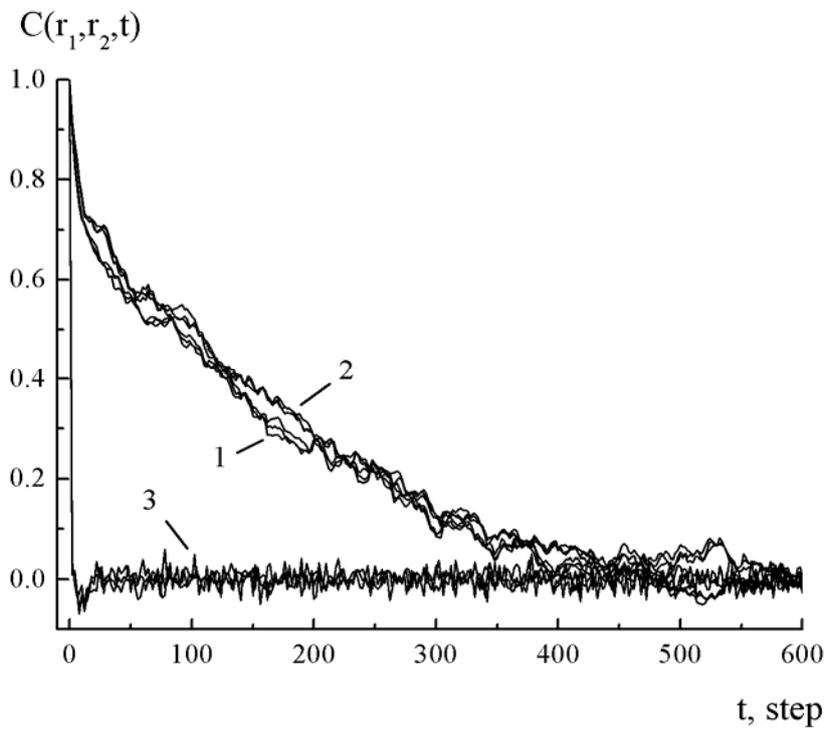

**Fig. 3.** Spatial and temporal spin correlation functions for the spin projections $S_x$, $S_y$, and $S_z$ in the directions of the $Ox$, $Oy$, and $Oz$ coordinate axes ($r_1 = r_2$).

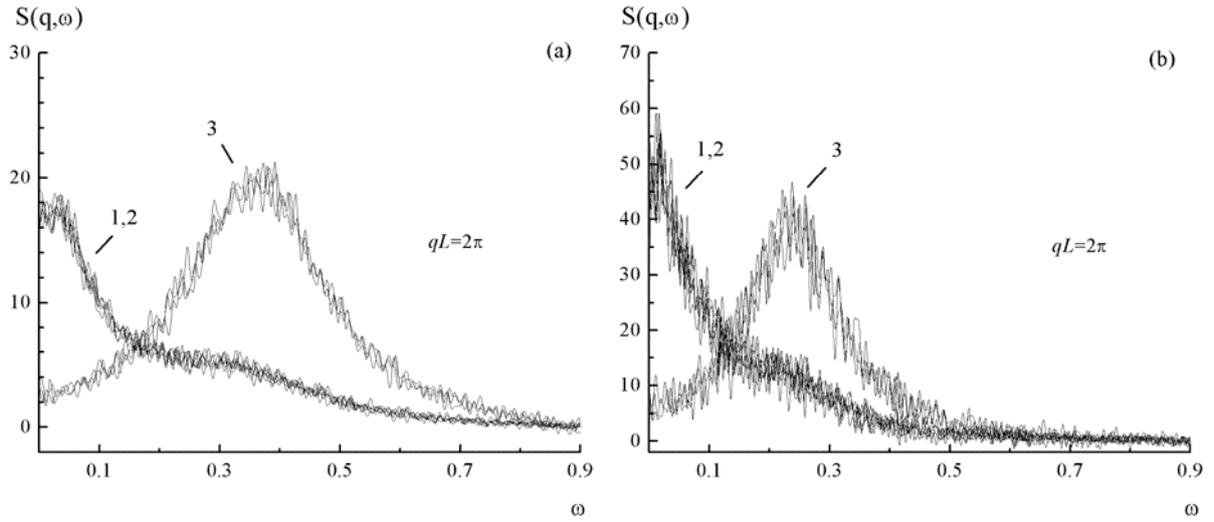

**Fig. 4.** Dynamic structure factors for the spin projections $S_x$, $S_y$, and $S_z$ in the directions of the $Ox$, $Oy$, and $Oz$ coordinate axes for the systems including (a) 1000 and (b) 2744 spins.

Figure 3 shows the time dependences of the space-time spin correlation functions $C(r_1, r_2, t)$, derived from the molecular dynamics equations for the system including 1728 spins, with $r_1 = r_2$. The correlations for three spin projections were considered in the directions of three coordinate axes. As can be seen from the graph, the correlations both between the $x$ projections and between $y$ projections in all three directions gradually tend to zero with time, whereas the correlations between the $z$ projections of the spins in all directions almost instantly drop to zero. This fact suggests the lack of correlation between the spin $z$ components.

Figure 4 presents the frequency dependences of the dynamic structure factors. These dependences are obtained by the Fourier transform of the space-time spin correlation functions over space and time. The data are given for the two systems with numbers of spins of 1000 (a) and 2744 (b). The considerable oscillations in the graphs (especially for the system with 2744 spins) are caused by the insufficient statistics on the initial equilibrium spin configurations. The numbers of such averagings are 80 and 17 for the systems with 1000 and 2744 spins, respectively.

It can be seen in Fig. 4 that the structure factors for the spin $x$ and $y$ projections in all three directions nearly coincide and differ from the structure factors for the spin $z$ projections (in the three directions as well). It can also be seen that the structure factor maxima increase with an increase in the size of the system.

Characteristic frequencies calculated from the three spin projections ($S_x$, $S_y$, and $S_z$) in the directions of the $Ox$, $Oy$, and $Oz$ coordinate axes (the data for two systems with $N = 1000$ and 2744 spins)

|       | $Ox$      | $Oy$      | $Oz$      |
|-------|-----------|-----------|-----------|
|       | $N=1000$  |           |           |
| $S_x$ | 0.149(6)  | 0.149(6)  | 0.147(6)  |
| $S_y$ | 0.145(6)  | 0.149(6)  | 0.148(6)  |
| $S_z$ | 0.358(6)  | 0.357(6)  | 0.359(6)  |
|       | $N=2744$  |           |           |
| $S_x$ | 0.028(6)  | 0.028(6)  | 0.028(6)  |
| $S_y$ | 0.028(6)  | 0.029(6)  | 0.027(6)  |
| $S_z$ | 0.153(6)  | 0.150(6)  | 0.158(6)  |

The characteristic frequencies $\omega_c$, calculated from the dynamic structure factors, are listed in the table. It can be seen that the $\omega_c$ values are the same for the spin $x$ and $y$ projections and differ from those obtained for the spin $z$ projections in each system, as in the case of structure factors. The characteristic frequency decreases within an increase in the size of the system.

Thus, the Heisenberg model with strong easy plane anisotropy describes fairly well both the static and dynamic critical behavior of the 3D *XY* model and its different types. Hende, the model considered here can be used to study the critical dynamics of magnetic superlattices.

REFERENCES

1. B. Hjörvarsson, J.A. Dura, P. Isberg, T. Watanabe, et al., Phys. Rev. Lett. **79**, 901 (1997).
2. V. Leiner, K. Westerholt, A.M. Blixt, H. Zabel, B. Hjörvarsson, Phys. Rev. Lett. **91**, 37202 (2003).
3. V. Leiner, K. Westerholt, B. Hjörvarsson, H. Zabel, J. Phys. D: Appl. Phys. **35**, 2377 (2002).
4. K.Sh. Khizriev, A.K. Murtazaev, V.M. Uzdin, J. Magn. Magn. Mater. **300**, e546 (2006).
5. A.K. Murtazaev, K.Sh. Khizriev, V.M. Uzdin, Bulletin of the Russian Academy of Sciences: Physics **70**, 695 (2006).
6. D.P. Landau, M. Krech, J. Phys.: Condens. Matter. **11**, R179 (1999).
7. M. Krech, D.P. Landau, Phys. Rev. B **60**, 3375 (1999).
8. A.K. Murtazaev, V.A. Mutailamov, I.K. Kamilov, et al., J. Magn. Magn. Mater. **258-259**, 48 (2003).
9. A.K. Murtazaev, V.A. Mutailamov, K.Sh. Khizriev, Bulletin of the Russian Academy of Sciences: Physics **68**, 835 (2004).
10. A.K. Murtazaev, V.A. Mutailamov, Journal of Experimental and Theoretical Physics **101**, 299 (2005).